\documentclass[12pt]{article}

\usepackage{amsmath, amssymb, amsthm}

\usepackage{graphicx}
\usepackage{caption}
\usepackage{float}

\usepackage[utf8]{inputenc}
\usepackage[greek,english]{babel}

\usepackage{geometry}
\usepackage[colorlinks=true, linkcolor=blue, citecolor=blue, urlcolor=blue]{hyperref}
\usepackage{bookmark}

\usepackage[most]{tcolorbox}
\tcbuselibrary{theorems}
\newtcbtheorem{mydefinition}{Definition}%
{colback=blue!5!white, colframe=blue!50!black, fonttitle=\bfseries,
arc=4mm, boxrule=0.8pt}{def}

\title{Comment on “Unifying Aspects of Generalized Calculus”}
\author{
\large Mikołaj Sienicki\thanks{Polish-Japanese Academy of Information Technology, ul. Koszykowa 86, 02-008 Warsaw, Poland, European Union.} 
\quad and \quad
Krzysztof Sienicki\thanks{Chair of Theoretical Physics of Naturally Intelligent Systems ($\mathbb{NIS}^{\text{\textcopyright{}}}$), Lipowa 2/Topolowa 19, 05-807 Podkowa Leśna, Poland, European Union.}
}
\date{\today}

\begin{document}

\maketitle

\begin{abstract}
\noindent Czachor’s recent proposal introduces a form of non-Newtonian calculus built by pulling back arithmetic operations through arbitrary bijections between continua. Although the idea is mathematically inventive, it runs into serious conceptual trouble when examined from a physical standpoint. Claims of universal applicability quickly unravel under scrutiny—especially when considering pathological bijections like the Cantor function, which break the framework’s core assumptions. When applied to domains such as relativity, entropy, or cosmology, the results often collapse into tautological restatements lacking real predictive power. This commentary explores these issues in depth, highlighting where and why the formalism falls short of providing a physically coherent theory.
\end{abstract}

\section*{Comment}

Czachor’s framework \cite{Czachor2020} (see also \cite{czachor2016relativity}) begins with a seemingly simple but bold idea: use bijections between sets of continuum cardinality to redefine arithmetic operations. Specifically, given bijections \(f_X: X \to \mathbb{R}\) and \(f_Y: Y \to \mathbb{R}\), one can pull back standard addition and multiplication from the real numbers to these sets. Formally speaking, the move is valid. But it opens the door to a serious problem: ambiguity. There are uncountably many such bijections between continuum-sized sets, and without a physical principle to narrow the field, the calculus turns arbitrary by construction.

And that arbitrariness isn't just a technical quirk—it carries real physical implications. Czachor’s framework assumes differentiability for these bijections, yet this doesn’t always hold. A striking counterexample is the Cantor function \(C(x)\): it's continuous and strictly increasing over \([0,1]\), yet its derivative is zero almost everywhere—and undefined on a dense set. Try plugging this into Czachor’s machinery: even a basic function like \(A(x) = x\) gives rise to a non-Newtonian derivative of the form

\begin{equation}
\frac{DA(x)}{Dx} = f_Y^{-1}\left( \frac{dA\circ C(x)}{dC(x)} \right),
\end{equation}

\noindent but this expression is undefined in most places. The derivative simply doesn't converge, since tiny changes in \(x\) can result in jumps in \(f_X(x)\). Czachor’s Eq.~(21), which claims \(Df_X(x)/Dx = 1\), doesn’t survive in the face of this pathology. Without differentiability, the entire structure loses its supposed universality.

\begin{figure}[h!]
    \centering
    \includegraphics[width=0.65\textwidth]{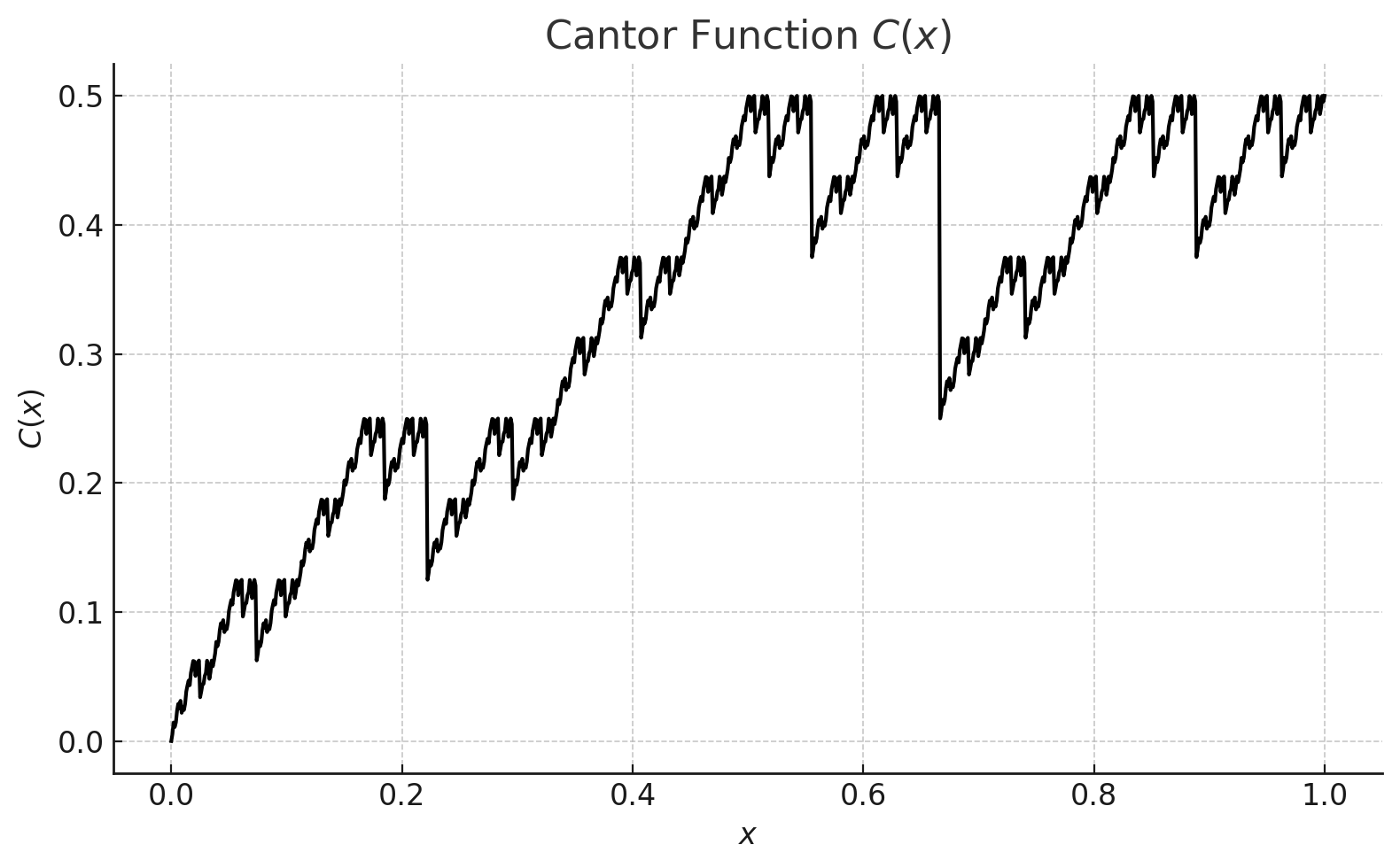}
    \caption{The Cantor function \( C(x) \), used here to illustrate a breakdown in the assumptions underlying non-Newtonian calculus. Although continuous and strictly increasing on \([0,1]\), it is constant on large intervals, satisfies \( C'(x) = 0 \) almost everywhere, and is nondifferentiable on a dense set. The non-Newtonian derivative \(\frac{DA(x)}{Dx} = f_Y^{-1}\left( \frac{dA \circ C(x)}{dC(x)} \right)\) becomes ill-defined across most of the domain.}
    \label{fig:cantor}
\end{figure}

This breakdown becomes even more apparent when one tries to apply the calculus to physical laws. Consider velocity addition. If you choose the bijection \(f_X(\beta) = \operatorname{arctanh}(\beta)\), you recover Einstein’s familiar relativistic formula:

\begin{equation}
\beta_1 \oplus_X \beta_2 = \frac{\beta_1 + \beta_2}{1 + \beta_1 \beta_2}.
\end{equation}

So far, so good. But now try \(f_X(\beta) = \beta^3\), which is smooth and bijective over \([-1,1]\)\footnote{The function \( f_X(\beta) = \beta^n \) is bijective and real-valued on the interval \([-1,1]\) only when \(n\) is an odd positive integer. For even integers, \(f_X\) is not injective on \([-1,1]\), as both \( \beta \) and \( -\beta \) map to the same value. For non-integer \(n\), \(f_X(\beta)\) becomes complex-valued for negative \( \beta \), making the inverse \( f_X^{-1} \) undefined on part of the domain. Therefore, to preserve the real-valued, invertible structure needed for non-Newtonian calculus across the full physical range of relativistic velocities (\(\beta \in [-1,1]\)), only odd integer values of \(n\) are admissible.
}
. The resulting formula becomes:
\begin{equation}
    \beta_1 \oplus_X \beta_2 = (\beta_1^3 + \beta_2^3)^{1/3}.
\end{equation}
\noindent Using \(\beta_1 = \beta_2 = 0.9\), this gives \(\beta \approx 1.133\)—a velocity greater than the speed of light. No one is suggesting this as a real physical law; that’s precisely the point. The cube function meets the mathematical criteria, yet the resulting physics is unambiguously wrong. Only one very specific bijection leads to the correct relativistic addition; the rest do not. The general framework allows physical laws to be redefined arbitrarily unless anchored by external constraints.

The choice of \( f_X(\beta) = \beta^3 \) offers a striking illustration of just how easily arbitrary bijections—while algebraically acceptable—can yield \textit{physically nonsensical outcomes}. The cube function, despite being smooth and bijective over the interval \([-1, 1]\), fails to respect one of the most basic empirical constraints: the relativistic speed limit. Under this choice, the velocity composition rule becomes
\begin{equation}
\beta_1 \oplus_X \beta_2 = \left( \beta_1^3 + \beta_2^3 \right)^{1/3},
\end{equation}
which, for example, gives \( \beta \approx 1.133 \) when both inputs are \( \beta_1 = \beta_2 = 0.9 \)—a superluminal result in clear conflict with special relativity. What this underscores is that the physical validity of predictions in the framework hinges entirely on how \( f_X \) is chosen.

In contrast, the bijection \( f_X(\beta) = \operatorname{arctanh}(\beta) \) is not arbitrary—it emerges naturally from Lorentz transformations and the geometry of Minkowski space. The cube function, by comparison, has no link to physical symmetries or structure. Its use not only breaks the speed limit, it also breaks the illusion of generality: without a guiding physical principle, the formalism permits dynamics that violate well-established empirical boundaries.

A similar breakdown appears when the formalism is applied to entropy. Czachor defines a non-Newtonian entropy function as
\begin{equation}
S_X(P) = f_X^{-1}\left( -\sum_i p_i f_X(\ln p_i) \right),
\end{equation}
\noindent which reduces to Shannon entropy when \( f_X(x) = x \). But when \( f_X(x) = e^x \), the formula transforms into
\begin{equation}
S_X(P) = \ln\left( -\sum_i p_i^2 \right),
\end{equation}
\noindent a quantity that is undefined for any legitimate probability distribution. Once again, whether the result makes sense physically—or makes sense at all—depends entirely on how the arithmetic has been redefined.

Further complications arise in Czachor’s use of \(\kappa\)-calculus, where alleged equivalences are drawn between expressions that only match under narrowly constrained assumptions. Likewise, the claim that cosmic acceleration can be explained without invoking dark energy falls apart under inspection. The proposed cosmological model turns out to be nothing more than a reparametrization of the standard \(\Lambda\)CDM solution, with the cosmological constant tucked away inside a carefully chosen bijection, \(f_X(t) \sim \sinh(\tfrac{3}{2}\sqrt{0.7}\,t)\). No new mechanism is proposed, and no additional predictive power is gained.

The treatment of Bell’s theorem in Czachor’s work \cite{czachor2020loophole,czachor2023contra} reveals a related problem. By redefining expectation values using non-Newtonian operations, the assumptions that underpin Bell's inequality are effectively sidestepped. This maneuver doesn’t amount to a genuine physical loophole—it’s a redefinition that renders the inequality inapplicable by design. At best, it's a semantic shift; at worst, it amounts to evasion rather than insight.

Czachor’s non-Newtonian calculus, for all its algebraic elegance, falls short of the basic demands of a physical theory. It offers no principle for choosing among the vast set of possible bijections, no way to test its predictions empirically, and no assurance of physical consistency. In the absence of such anchors, the framework reads more like a clever relabeling of conventional structures than a tool for understanding nature. Its supposed universality is undermined by the very freedom it affords—and with that, its scientific value remains unproven.

This critique is echoed in Lambare’s independent analysis \cite{lambare2021comment}, which takes direct aim at Czachor’s claim that non-Newtonian arithmetic opens a loophole in Bell-type theorems. While acknowledging the internal consistency of the mathematics, Lambare emphasizes a point central to our argument: the formalism does not preserve essential physical assumptions.

In particular, Lambare demonstrates that Czachor’s modified expectation values take the form
\begin{equation}
E(A, B) = \int d\lambda\, \rho(\lambda \mid A, B)\, A(a, \lambda) \odot B(b, \lambda),
\end{equation}
\noindent nowhere \( \odot \) denotes a non-Newtonian product and \( \rho(\lambda \mid A, B) \) is a distribution that depends explicitly on the measurement settings. This violates the principle of \emph{measurement independence}—a cornerstone of Bell’s theorem. As a result, the violation of the inequality becomes trivial: it occurs not because the theorem has been refuted, but because its preconditions have been abandoned.

This reflects a deeper problem running through Czachor’s framework. By allowing arbitrary bijections to redefine fundamental operations, the formalism permits symbolic transformations that lack physical justification. Whether applied to velocity addition, entropy, or quantum correlations, the approach produces either contradictions or meaningless results—unless external physical constraints are imposed. Just as \( f_X(\beta) = \beta^3 \) leads to superluminal predictions, redefining probability through exotic arithmetic leads to violations of quantum limits—but not because nature demands it, only because the notation allows it.

Lambare’s conclusion is clear: Czachor’s model doesn’t overturn Bell’s theorem; it simply reframes the question in a way that avoids its premises. And that, in turn, supports our broader critique. This is not a discovery of new physics—it is a reshuffling of mathematical language that risks obscuring rather than illuminating physical reality. Whether in the classical or quantum domain, Czachor’s calculus edges dangerously close to being mathematically polished but scientifically inert.

To step back: Marek Czachor’s non-Newtonian arithmetic is, without question, a bold and imaginative proposal—an attempt to retool the language of physics by way of algebraic generalization. Yet despite its internal consistency, the theory remains untethered from empirical mechanisms. The central idea—redefining arithmetic operations via pullbacks through arbitrary bijections—is formally clever but physically directionless.

The framework permits the construction of dynamics that are mathematically rigorous and physically incoherent in equal measure, depending wholly on the arbitrary choice of bijection. Its applications to relativity, thermodynamics, cosmology, and quantum foundations share a common trait: they replace explanatory substance with symbolic substitution. And in none of these domains does the formalism yield new testable predictions or resolve outstanding puzzles. If anything, it pushes those puzzles out of view by altering the terms of the discussion.

In short, Czachor’s work offers an intellectually stimulating algebraic exercise—but in its current form, it lacks the structure needed to become a meaningful physical theory. Without empirical constraints, predictive insight, or interpretive clarity, it risks becoming a sophisticated system of relabeling, rather than a path toward deeper understanding.

A more cautiously balanced view comes from Piotr Piłat, whose recent paper, \textit{Bell-Type Inequalities from the Perspective of Non-Newtonian Calculus—A Critical Reappraisal} \cite{pilat2024}, offers a tempered take on the issues at hand. While Piłat echoes many of the mathematical concerns we’ve raised, he approaches their implications with a lighter touch.

He acknowledges that Czachor’s use of non-Newtonian arithmetic significantly alters the foundational mathematical structures—most notably, the expectation values that underpin Bell-type inequalities. In doing so, the formalism enables violations of those inequalities, but only in a symbolic sense. Piłat concedes that this represents a shift in assumptions rather than a breach of Bell’s theorem itself. On this point, his reading aligns with ours: what’s at stake is a semantic maneuver, not an empirical challenge.

Where Piłat diverges is in tone and emphasis. He refrains from labeling Czachor’s framework scientifically hollow. Instead, he leaves room for the idea that alternative arithmetics might shed new light on quantum correlations—so long as their interpretive departure from conventional physics is made explicit. Unlike our critique, which sees the absence of empirical constraint as disqualifying, Piłat treats the formalism as potentially valuable within a more philosophical or representational context.

In this light, Piłat’s contribution acts as a moderating voice. He does not defend the physical adequacy of non-Newtonian models, but he also stops short of dismissing them outright. Rather, he positions them as symbolic reformulations worth exploring—provided they are understood as such. His analysis ultimately reinforces our core concern: Czachor’s framework, as it stands, lacks the physical grounding needed to function as a viable theory. Yet Piłat also cautions against ruling out its broader interpretative potential too hastily.

A recent mathematical study by Torres \cite{Torres2023} presents a non-Newtonian version of the calculus of variations, extending Noether’s theorem into the multiplicative or generalized arithmetic setting. While this work is mathematically consistent and intellectually stimulating, it does not invalidate the critiques raised in our present article.

Most importantly, Torres does not claim that standard physical laws are incorrect, nor does he propose empirical consequences arising from his generalized framework. His work remains within the domain of formal generalization. As such, it lacks any assertion that the modified calculus has direct physical relevance. This is a crucial distinction: while Torres shows that a Noether-like structure can be preserved under redefined arithmetic, he does not address whether standard conservation laws or physical symmetries remain intact under real-world conditions.

Our critique remains focused on the misuse of non-Newtonian arithmetic in physical modeling. The danger arises when mathematical constructions, such as those discussed in \cite{Torres2023}, are prematurely interpreted as revealing hidden physical truths without empirical justification. The lesson is clear: \textbf{mathematical generality does not imply physical necessity}.

\bigskip

\noindent Our analysis critically challenges the use of such formalisms (e.g., by Czachor) to argue that:

\begin{itemize}
  \item “the laws of physics must be rewritten,”\footnote{Czachor does not explicitly state that “the laws of physics must be rewritten,” but he strongly implies this view. In Section 5.1 of \cite{Czachor2020}, he writes: “A reformulation of physics may become necessary if [the assumption that arithmetic is absolute] is relaxed.” Similarly, in the introduction he challenges the standard assumption that physical arithmetic must be Diophantine: “What if this assumption is wrong?” These statements justify our interpretative paraphrase.}

  \item “standard arithmetic is merely a choice,”
  \item “Bell and Einstein were wrong because they used the wrong arithmetic.”
\end{itemize}

\noindent We demonstrate that:

\begin{itemize}
  \item the arbitrariness of bijections \( f_X \) makes the theory unfalsifiable,
  \item one can always choose a bijection to ‘predict’ anything — which undermines the scientific value of the formalism,
  \item Czachor’s results are not “new laws of physics,” but symbolic reformulations devoid of new empirical content.
\end{itemize}

Czachor’s non-Newtonian arithmetic provides a thought-provoking extension of mathematical formalism. But its physical applications require a firm empirical foundation. We caution against mistaking symbolic flexibility for physical generality. Future developments must focus on selecting bijections based on symmetry, invariance, and observational relevance.

Until such grounding is provided, the framework remains mathematically elegant but physically inconclusive.

\bigskip
\textbf{Declaration of Interests:} The authors declare no conflicts of interest. Parts of this appendix were refined using AI tools such as ChatGPT, Gemini, Copilot, and Grammarly for translation and editing assistance. All content, however, remains the responsibility of the authors.

\appendix
\section*{Appendix: On Reciprocal Means, Cauchy Additivity, and Physical Misinterpretations}

A follow-up analysis by the authors \cite{Sienicki2025} takes a closer look at the broader implications of Czachor’s non-Diophantine framework, especially how it has been applied—or, as we argue, misapplied—to fundamental laws in physics. Czachor suggests that even well-established linear laws such as Ohm’s Law, Kirchhoff’s rules, or average velocity formulas point toward a need for a generalized arithmetic based on non-linear bijections \cite{Czachor2020, Czachor2016}. But a more careful examination reveals that these claims rest on misunderstandings of standard physical reasoning.

\paragraph{1. The Harmonic Mean is Not Evidence of New Arithmetic.}  
Czachor points to the harmonic mean—used, for instance, in calculating average speeds or equivalent resistances in parallel—as evidence of an underlying "non-standard" arithmetic \cite{Czachor2020}. But as we’ve shown \cite{Sienicki2025}, this use of the harmonic mean naturally emerges from linear relationships, not from any departure from ordinary arithmetic. The formula for resistors in parallel, for example, follows directly from current conservation and Ohm’s Law. The suggestion that these cases require redefined arithmetic is, in our view, a misinterpretation of how reciprocal operations work—not a valid challenge to Diophantine arithmetic.

\paragraph{2. The Breakdown of Cauchy Additivity.}  
Another problem is mathematical. If we define a new addition rule using a non-linear bijection—say \( f_X(x) = \tanh(x) \)—then the new addition operation,

\begin{equation}
f_X(x + y) = f_X(x) \oplus f_X(y),    
\end{equation}

violates the Cauchy functional equation, which only holds if \( f_X \) is affine. As discussed in \cite{Sienicki2025}, this breakdown undercuts the interpretation of \( \oplus_X \) as a meaningful composition law in physics. We’ve already shown earlier that choosing bijections like \( f_X(\beta) = \beta^3 \) leads to absurd results—such as superluminal velocities—which suggests that the mathematics is not guiding physics here, but rather distorting it \cite{Czachor2020}.

\paragraph{3. The Concept of Average Speed Requires No New Mathematics (A High School Example)}

Consider the example of a vehicle (\textit{sic})traveling at two different speeds over equal distances. The correct average speed is given by the harmonic mean:

\begin{equation}
\bar{v} = \frac{2 v_1 v_2}{v_1 + v_2}.
\end{equation}

This is derived from basic definitions: total distance divided by total time. Nothing about this calculation requires us to abandon standard arithmetic. Recasting it using non-Newtonian operations, as Czachor proposes, doesn’t enhance our understanding—it just adds symbolic overhead that, in the end, tells us nothing new \cite{Sienicki2025}.

\paragraph{4. What’s the Real Issue?}  
At heart, the problem is conceptual. Czachor’s mathematical formalism is certainly clever, but when applied to physics, it lacks a mechanism for deciding which bijection \( f_X \) to use. Without a physical principle to guide that choice, any law can be “explained” retroactively by tweaking \( f_X \). That’s not scientific explanation; that’s symbolic relabeling. The same vulnerability shows up whether we're talking about entropy, relativistic velocity, or Bell-type inequalities \cite{Sienicki2025}.

In short, this appendix strengthens the core message of our main text: non-Newtonian arithmetic, while mathematically interesting, does not in itself yield new physical insight. For a mathematical framework to contribute to physics, it must do more than offer formal flexibility—it must be grounded in empirical relevance and capable of making testable predictions.


\begin{thebibliography}{99}

\bibitem{Czachor2020}
Czachor, Marek. ``Unifying Aspects of Generalized Calculus.'' \textit{Entropy} 22, no. 11 (2020): 1180. \url{https://www.mdpi.com/1099-4300/22/10/1180}

\bibitem{czachor2016relativity}
Czachor, Marek. "Relativity of arithmetic as a fundamental symmetry of physics." Quantum studies: mathematics and foundations 3, no. 2 (2016): 123-133.
\url{https://link.springer.com/content/pdf/10.1007/s40509-015-0056-4.pdf}.

\bibitem{czachor2020loophole}Czachor, Marek. "A loophole of all ‘loophole-free’Bell-type theorems." Foundations of Science 25, no. 4 (2020): 971-985.\url{https://link.springer.com/content/pdf/10.1007/s10699-020-09666-0.pdf}

\bibitem{czachor2023contra} Czachor, Marek. "Contra Bellum: Bell's theorem as a confusion of languages." arXiv preprint arXiv:2301.10727 (2023).\url{https://arxiv.org/pdf/2301.10727}

\bibitem{lambare2021comment}Lambare, Justo Pastor. "Comment on “a loophole of all “loophole-free” Bell-type theorems”." Foundations of Science 26, no. 4 (2021): 917-924. \url{https://arxiv.org/pdf/2008.00369} see also Czachor, Marek. "Response to Comment on" A Loophole of All" Loophole-Free" Bell-Type Theorems", by JP Lambare." arXiv preprint arXiv:2008.11910 (2020).\url{https://arxiv.org/pdf/2008.11910}

\bibitem{pilat2024}
Piłat, Michał Piotr. "Bell-type inequalities from the perspective of Non-Newtonian calculus." Foundations of Science 29, no. 2 (2024): 441-457.
\url{https://link.springer.com/content/pdf/10.1007/s10699-022-09866-w.pdf}

\bibitem{Czachor2016}
Czachor, M. (2016). Relativity of Arithmetic as a Fundamental Symmetry of Physics. \textit{Quantum Studies: Mathematics and Foundations}, 3(2), 123–133. \url{https://doi.org/10.1007/s40509-015-0056-4}

\bibitem{Sienicki2025}
Sienicki, M., \& Sienicki, K. (2025). A Critical Analysis of Non-Diophantine Arithmetic and its Misinterpretations in Physics. Unpublished manuscript.

\bibitem{Torres2023}
Torres, Delfim FM. "A non-Newtonian Noether's symmetry theorem." Applicable Analysis 102, no. 7 (2023): 1934-1941.\url{https://arxiv.org/pdf/2111.11559}


\end{thebibliography}
\end{document}